\documentclass[11pt]{article}
\usepackage[preprint]{acl}
\usepackage{times,latexsym}
\usepackage[T1]{fontenc}
\usepackage[utf8]{inputenc}
\usepackage{microtype}
\usepackage{graphicx,booktabs,amsmath,amssymb}
\newcommand{\QueryCount}{189}
\newcommand{\PoolCount}{3,605}
\newcommand{\TestFamilies}{197}
\newcommand{\GraphPFive}{77.2}
\newcommand{\SequencePFive}{58.8}
\newcommand{\PairedGain}{18.4}
\newcommand{\GainLow}{15.0}
\newcommand{\GainHigh}{22.1}
\newcommand{\SearchMedian}{44}
\newcommand{\SearchTail}{323}
\newcommand{\AlignMedian}{415}
\newcommand{\AlignTail}{504}
\newcommand{\CandidateSuccess}{100.0}
\newcommand{\CandidateRecall}{63.1}
\newcommand{\ControlCount}{197}

\newcommand{\corpusRows}{%
Molweni & 10,000 & 88,303 & 13,033 \\
Reddit discourse & 9,483 & 115,827 & 19,157 \\
CGA--CMV & 6,842 & 42,964 & 6,466 \\
DeliData & 500 & 14,003 & 2,296 \\
WildChat shard & 36,821 & 217,432 & 32,454 \\
PRISM & 8,011 & 53,646 & 9,530 \\
DailyDialog train & 11,118 & 87,170 & 14,361 \\
Chromium & 1,484,843 & 2,853,498 & 54,192 \\
}

\newcommand{\retrievalRows}{%
TF--IDF (text) & 1.0 & 1.0 & 9.5 \\
MiniLM (text) & 1.2 & 1.1 & 9.5 \\
Sequence + speaker & 58.8 & 52.9 & 94.7 \\
+ target correspondence & 77.2 & 65.9 & 98.4 \\
}

\graphicspath{{../figures/}}
\hypersetup{pdftitle={Conversational DNA: A Visual Language and Interactive Atlas of Human and AI Dialogue},pdfauthor={Baihan Lin},pdfkeywords={dialogue visualization, visual analytics, conversational dynamics, multiparty dialogue, conversation analysis, human-AI interaction, discourse structure, conversation atlas}}
\title{Conversational DNA: A Visual Language and Interactive Atlas\\of Human and AI Dialogue}
\author{Baihan Lin \\
Berkman Klein Center for Internet \& Society, Harvard University \\
\texttt{blin@cyber.harvard.edu}}
\begin{document}
\maketitle
\begin{abstract}
What makes a conversation hold together when its participants speak across one another? Topic maps offer one view, but they leave the relationships between contributions difficult to inspect. We present Conversational DNA, a visual language and interactive atlas for exploring human and AI dialogue. Speaker strands preserve participation, communicative bases mark moves, and directed pairings connect responses to their targets. Adjustable helix geometry makes speaker switching, response distance, and contribution length visible. Across eight corpora containing 1.57 million source records, the atlas maps 151,489 indexed episodes and connects cohort comparison to source transcripts, local structural alignment, and recorded reply alternatives. On \QueryCount{} held-out Molweni motif queries, adding target correspondence improves precision@5 from \SequencePFive\% to \GraphPFive\% for exact annotated structure. Case readings illustrate interleaved participation, delayed responses, and the influence of annotation coverage on apparent collection differences. The system supports a view of conversation as jointly organized activity, with visual patterns serving as starting points for examining evidence rather than substitutes for interpretation.\footnote{Demo video: \url{https://youtu.be/S4SMteXzJb0}; Git repository for reproduciblity: \url{https://github.com/doerlbh/ConversationalDNA}}
\end{abstract}

\section{Introduction}
An answer can address an earlier question while another conversational branch continues. A clarification can itself prompt a clarification. In group conversation, the order of messages and the order of responses need not coincide. A useful visual account must therefore connect \emph{sequence}, \emph{participation}, and \emph{response attachment} without collapsing them into one alternating transcript.

This is also a linguistic distinction. Turn taking organizes opportunities to contribute \citep{sacks1974turntaking}, while repair concerns how participants address trouble in speaking, hearing, and understanding \citep{schegloff1977repair}. A discourse label offers a partial observation of that organization, not a complete explanation. An annotated answer may be unhelpful; a missing link may reflect missing annotation. Making these distinctions visible is valuable for dialogue researchers, conversation analysts, and developers auditing multi-turn AI behavior.

\begin{figure*}[t]
\centering\includegraphics[width=\textwidth]{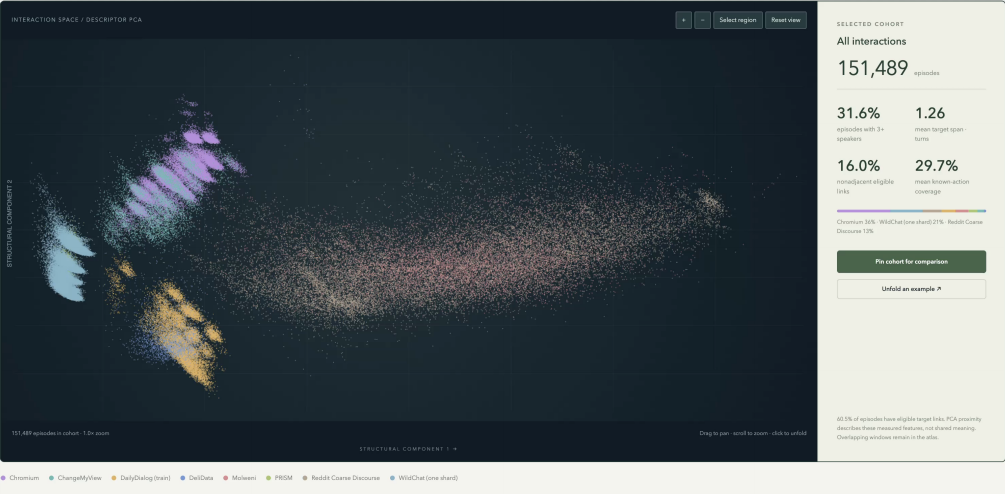}
\caption{The full episode atlas, captured from the running application. Every plotted point corresponds to one of 151,489 indexed episodes; clicking unfolds its conversation. Colors distinguish eight source corpora. Filters and region selection update the cohort panel, including participation, target span, and annotation coverage. The two principal components retain 44.4\% of variance in 19 standardized descriptors. Proximity is structural, not semantic; corpus separation can reflect annotation availability. Overlapping windows remain in the map.}
\label{fig:atlas}
\end{figure*}

Conversational DNA connects an explicit visual grammar to an interactive atlas (Figure~\ref{fig:atlas}). Users move from a collection-level pattern to a cohort, a response pairing, an individual exchange, and a local structural analog. The metaphor supports this movement across scales: \emph{bases} denote communicative moves, \emph{strands} track speakers, \emph{pairings} link responses to their targets, \emph{motifs} describe local configurations, and \emph{variants} compare recorded alternative replies. Each operation retains a path to source evidence.

We contribute a visual grammar with explicit linguistic and geometric meanings, an atlas that connects collection-level patterns to their source evidence, and a target-aware alignment method evaluated on annotated interaction motifs. Together, these let a reader ask where a pattern occurs, how its participants are connected, and whether the underlying words support the initial visual reading. The DNA metaphor provides a consistent way to navigate these questions across scales; it does not imply biological inheritance or a universal conversational alphabet.

\section{Related Work and Design Gap}
WildVis supports million-scale chat exploration with text and embedding search \citep{deng-etal-2024-wildvis}. Dynamics summaries abstract from conversational topic \citep{hua-etal-2024-get}, and ConDynS compares ordered patterns and transcripts \citep{jung-etal-2025-similarity}. OnGoal visualizes conversational goal progress in an interactive writing workflow \citep{coscia2025ongoal}. These systems motivate complementary questions about content, trajectory, and goals. Our design centers a different connected task: finding a local interaction pattern and inspecting its precise speaker and response-target correspondence. Our evaluation does not reproduce these systems' protocols or establish general superiority.

Discourse sequences \citep{zhang2017coarse}, multiparty dependencies \citep{li-etal-2020-molweni}, and conversation infrastructure \citep{chang-etal-2020-convokit} supply reusable foundations. DNA metaphors also predate this system in eye-movement visualization \citep{deitelhoff2019dna} and the analysis of visual-encoding building blocks \citep{engelhardt2020dna}. We claim neither the metaphor nor the helix as new. The design contribution is the explicit mapping from conversational evidence to visual channels, coupled to corpus-scale inspection and retrieval. Its perceptual benefits remain a testable hypothesis.

\section{A Linguistic Visual Language}
\subsection{Sequence, participation, and relevance}
The helix is a readable representation of an ordered dialogue graph (Figure~\ref{fig:readings}A). Vertical position preserves source turn order. Each speaker receives a continuous, colored strand, with a bead only where that speaker actually contributes. A strand maintains identity through intervening turns; its unoccupied segments do not imply speech. Thus the visual backbone separates participation from simple adjacency.

Each bead carries a short move code, such as \texttt{CLQ}, \texttt{ANS}, or \texttt{ACK}. These abbreviate clarification, answer, and acknowledgement; they are not biological nucleotides or an exhaustive theory of speech acts. Source-native labels remain available. In Molweni, several node labels are derived from annotated discourse relations, so they should not be mistaken for independently annotated intentions.

Directed pairings connect a response bead to its recorded target, including nonadjacent targets. This makes \emph{attachment} visible: which contribution the source annotation or platform says a turn addresses. Attachment is related to, but weaker than, the conversation-analytic concept of sequential relevance. Successful repair requires examining what participants do with the contribution \citep{schegloff1977repair}. The interface therefore links every selected bead to the original text and provenance.

\subsection{Geometry with declared meanings}
Three adjustable geometric channels complement the discrete structure (Table~\ref{tab:grammar}). Local speaker switching increases angular advance per turn, tightening the helix's participation rhythm. Response-target distance expands its radius, exposing contributions that reach farther through source order. Bead radius increases with the square root of whitespace-delimited word count, showing contribution length. None measures topic coherence, semantic distance, or linguistic complexity. Exact definitions and fixed-reference settings appear in Appendix~\ref{app:geometry}.

The visual-grammar workspace explains these channels beside an interactive specimen. Users can rotate its viewpoint, fix pitch or radius, hold bead size constant, and recolor beads by move. Speaker-colored backbones retain identity in either coloring mode. Depth shading supports the apparent helix geometry; it encodes no confidence or sentiment. Solid links indicate eligible annotations or platform targets, dashed links indicate chronology, and unknown actions remain explicit. The fixed-lane view and transcript provide complementary ways to verify an attractive but potentially ambiguous shape.

\subsection{From bases to motifs and variants}
A two-strand picture would obscure the organization of a group exchange. Our design therefore assigns a strand to each participant and a base to each contribution: moves belong to turns, not to people. Pairings can branch, and their endpoints need not be neighbors in the transcript. The helix supplies a shared visual backbone for these relationships; it does not impose biological complementarity on them. This choice has a cost. More participants create crossings, and apparent depth can complicate target tracing. Rotation, selected-link emphasis, fixed geometry, and the alternative lane view let readers check a reading across representations. Whether this combination helps people reason more accurately remains an empirical question.

\begin{figure*}[t]
\centering\includegraphics[width=\textwidth]{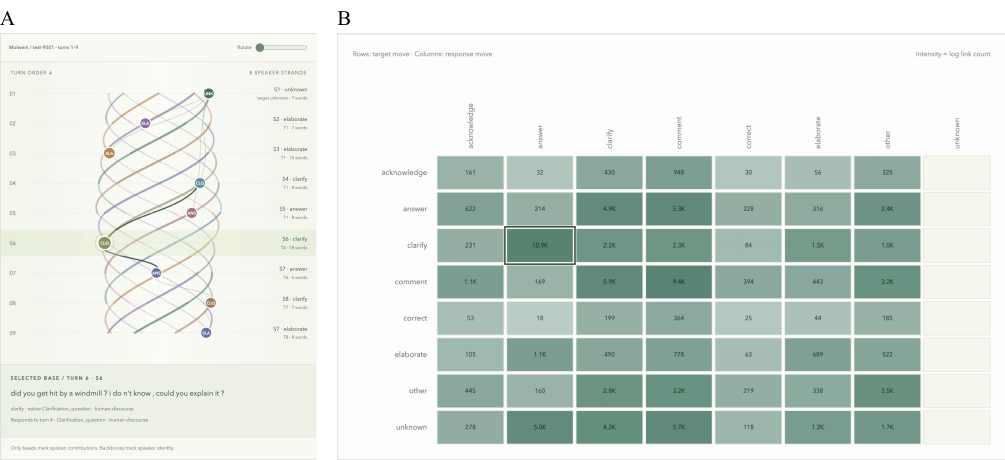}
\caption{Two linked readings from actual interface captures. (A) The visual-grammar specimen shows nine turns and eight speaker strands in Molweni \texttt{test-9001}. Selected turn 6 (\texttt{CLQ}) targets turn 4 across turn 5; turn 7 (\texttt{ANS}) targets turn 6. The text below the helix makes this annotation inspectable. (B) The Molweni target--response census contains 10,862 clarification-to-answer link occurrences across indexed windows. Selecting that cell opens an actual containing episode. These are overlapping-window counts, not unique pairs or verified instances of successful repair.}
\label{fig:readings}
\end{figure*}

A motif combines moves, speaker relationships, and response topology. Its identity is relational: an answer attached to the wrong clarification is a different pattern even if the label sequence is unchanged. The pairing census aggregates target-move to response-move links within one corpus; selecting a cell opens an episode containing that pair (Figure~\ref{fig:readings}B). PRISM's recorded reply alternatives support a separate notion of variation \citep{NEURIPS2024_be2e1b68}. Unchosen replies remain terminal alternatives, without invented downstream trajectories. This gives the metaphor concrete analysis operations while avoiding claims of mutation, inheritance, or evolutionary fitness.

\section{Atlas, Representation, and Retrieval}
\subsection{Collection and evidence model}
The research index contains eight source subsets (Table~\ref{tab:corpora}). Chromium contributes many short code-review records \citep{meyers-etal-2018-dataset}. Molweni, Reddit discourse, ChangeMyView \citep{chang-danescu-niculescu-mizil-2019-trouble}, and DeliData \citep{karadzhov2023delidata} provide group settings; DailyDialog contributes annotated acts \citep{li-etal-2017-dailydialog}. WildChat supplies human--AI interactions \citep{zhao2024wildchat,wildchat48card}, and PRISM supplies selected paths and recorded alternatives. This is a heterogeneous convenience collection, not a representative sample of humanity.

\begin{table}[t]
\centering\small\setlength{\tabcolsep}{3pt}
\begin{tabular}{lrrr}\toprule
Source subset & Records & Utterances & Episodes\\\midrule
\corpusRows
\midrule Total & 1,567,618 & 3,472,843 & 151,489\\\bottomrule
\end{tabular}
\caption{Actual indexed subsets. Records follow each source's conversation unit and retain source-level duplicates. Episodes are bounded windows drawn from 117,310 records; the atlas plots every indexed episode.}
\label{tab:corpora}
\end{table}

A conversation is an ordered graph with text, within-conversation speaker aliases, native and display labels, and annotation provenance. An edge $(i,j,r,p)$ means turn $i$ responds to target $j$ with relation $r$ and provenance $p$. Human discourse annotations and platform reply links are eligible for target analysis; chronological predecessors are not. Unannotated user/assistant roles remain roles. Labels across corpora are not assumed equivalent. Episode slicing preserves external targets as boundary metadata, and the ordinal axis denotes source order rather than elapsed time.

The index uses windows of at most 12 turns, stride six, minimum four turns, and at most 64 windows per record. The map describes all 151,489 resulting episodes using 19 measured features: participation, action coverage and composition, response span and branching, utterance length, and lexical recurrence. Standardization and two-component PCA retain 27.0\% and 17.4\% of variance. Loadings and scaling parameters are saved. Missing-target features use zero for projection and retain an availability indicator; the cohort panel instead displays unavailable target statistics as N/A. Small deterministic display offsets separate coincident points. This is an inspectable structural projection, not a learned semantic landscape.

\subsection{Interaction across scales}
The Canvas atlas supports pan, zoom, point inspection, region selection, corpus filtering, and lenses for multiparty, nonadjacent, clarification, and branching episodes. Pinning a cohort exposes differences in participation and annotation coverage as another region or filter is selected. Representative source episodes appear as miniature strands beneath the map. Opening an episode links its helix, transcript, native labels, and response targets. Cohort export records episode IDs, projection metadata, filters, and comparison statistics; episode export preserves source evidence and any in-memory annotation overlay.

\begin{figure*}[t]
\centering\includegraphics[width=\textwidth]{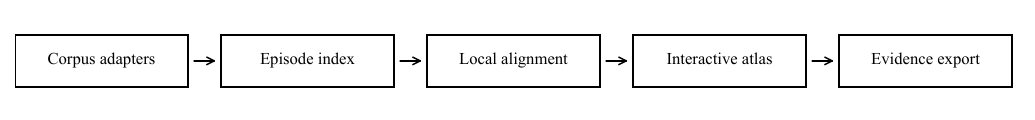}
\vspace{-2em}
\caption{Implementation components. Corpus adapters preserve source evidence; the episode index stores structural tokens and projection descriptors; local alignment retrieves correspondences; the interface links the atlas, DNA specimen, and source text; exports record the selected evidence and analysis settings. FastAPI serves SQLite data to Canvas and SVG views. Arrows indicate the analysis workflow, not a strictly linear runtime dependency.}
\label{fig:architecture}
\end{figure*}

\subsection{Target-aware local alignment}
SQLite FTS5 proposes up to 120 candidates using labels, adjacent label pairs, speaker transitions, and typed-edge tokens. A Smith--Waterman-style dynamic program \citep{smith1981} proposes up to 12 monotone local alignments per candidate. The score combines normalized label alignment $A$, pairwise speaker-identity agreement $S$, and typed-edge correspondence $F_E$:
\begin{equation}
R(M)=0.50A(M)+0.15S(M)+0.35F_E(M).
\label{eq:score}
\end{equation}
An edge matches only if both endpoints and its native relation type correspond. When neither graph contributes an eligible edge, the remaining weights are renormalized. The highest-scoring proposed alignment represents the candidate. These are fixed engineering settings, not fitted test-set parameters; the method is approximate, not optimal graph matching. The interface exposes aligned text, target correspondence, and coverage. Appendix~\ref{app:alignment} supplies the full scoring definitions.

\section{Atlas Readings and Validation}
\subsection{What the visual language reveals}
\paragraph{Interleaving and repair candidates.}
In Figure~\ref{fig:readings}A, eight participants contribute within nine turns. The highlighted clarification addresses turn 4 while turn 5 responds elsewhere. The subsequent answer targets the clarification. Following beads and pairings distinguishes these relationships from simple alternation. Reading the source also reveals an informal, partly ambiguous exchange: the labels support inspecting a repair candidate, not asserting that understanding was achieved. This is a case reading of an existing annotation, not an independent discourse analysis.

\paragraph{Collection structure and missingness.}
The Molweni nonadjacent-response lens retains 11,533 episodes: 76.1\% contain at least three speakers, compared with 31.6\% across the full map. Their mean known-action coverage is 89.9\%, compared with 29.7\% globally (Appendix~\ref{app:screens}). The interface shows these differences together. The largest positive PC1 loadings are nonadjacent-link fraction (0.379) and target branching (0.366), but known-action coverage also contributes (0.309). Otherwise, a dense region of annotated human dialogue could be misread as inherently richer than sparsely annotated AI dialogue. These descriptive contrasts concern indexed windows and selection policies, not causal or population differences.

\paragraph{A reason to preserve targets.}
Across the original Molweni conversations, 27,462 of 78,245 annotated links (35.1\%) join nonadjacent utterances. Pairings retain these dependencies. The census in Figure~\ref{fig:readings}B offers a complementary aggregate view, with explicit overlap and provenance qualifications. The original-link count and window-level census use different denominators.

\subsection{Known-answer structural retrieval}
We use Molweni's official train/test split with additional family separation based on shared normalized utterances. Connected components group overlapping records; we retain one training representative per family, exclude train-connected test families, and retain one representative per remaining test family. This yields \PoolCount{} training candidates and \TestFamilies{} test families. Each eligible test family supplies one ordered three-node, two-edge query motif with 2--100 relevant training records, producing \QueryCount{} queries.

An independent exhaustive enumerator defines relevance by matching labels, pairwise speaker identity, directed topology, and native edge types. The two structural conditions share the same 120 candidates; the principal ablation removes $F_E$ and renormalizes the remaining weights. TF--IDF and MiniLM \citep{reimers-gurevych-2019-sentence,minilmcard} rank the full training pool by text, receiving less structural information. They are reference points for this exact-structure task, not implementations of WildVis or ConDynS.

\begin{table}[t]
\centering\small\setlength{\tabcolsep}{3pt}
\begin{tabular}{lrrr}\toprule
Ranking method & P@5 & P@10 & Success@10\\\midrule
\retrievalRows
\bottomrule\end{tabular}
\caption{Exact annotated motif retrieval, percentages, over \QueryCount{} family-separated queries. Success@10 indicates at least one relevant result. The bottom two rows share candidates and annotations. This evaluates structural retrieval, not semantic relevance or visual comprehension.}
\label{tab:retrieval}
\end{table}

Target correspondence increases P@5 from \SequencePFive\% to \GraphPFive\% (Table~\ref{tab:retrieval}). The paired gain is \PairedGain{} percentage points, with 95\% bootstrap interval [\GainLow, \GainHigh] over 2,000 query-family resamples. Candidate generation retains a positive for \CandidateSuccess\% of queries but only \CandidateRecall\% mean recall over all positives. It remains a bottleneck. In a separate control, rerouting one eligible edge in each of \ControlCount{} retained test-family conversations leaves transcripts and labels unchanged. Target-aware alignment prefers the original every time; the sequence ablation always ties. This confirms sensitivity to supplied structure, not superiority over text metrics with identical inputs.

\subsection{Implementation and operational checks}
Python adapters, SQLite, FastAPI, Canvas, and SVG implement the local system without an inference API key. The record index occupies approximately 2.6\,GiB; a separately built sidecar supplies all episode coordinates and descriptors. On local ARM64 macOS, warm HTTP text search has median/p95 latency \SearchMedian/\SearchTail\,ms (30 requests); matching has \AlignMedian/\AlignTail\,ms (40 requests, five corpora). These exclude browser rendering and hosted-network effects. Automated checks cover speaker invariance, boundary links, target sensitivity, chronology exclusion, immutable overlays, and atlas-to-source agreement. Browser checks exercise cohort filtering, DNA selection, the pairing-to-episode workflow, and alignment inspection. They do not establish perceptual effectiveness.

The installable companion package includes the service, interface, scripts, results, and clearly labeled synthetic examples. Full-corpus mode obtains sources separately under their own terms. New code uses the MIT license. A narrated walkthrough demonstrates the connected workflow; access details must accompany the submitted package and video.

\begin{table}[t]
\centering\small\setlength{\tabcolsep}{4pt}
\begin{tabular}{p{.25\columnwidth}p{.65\columnwidth}}
\toprule Visual channel & Linguistic reading and measured quantity\\\midrule
Axial order & Sequence: source position, not elapsed time.\\[3pt]
Speaker strand & Participation: identity persists through others' turns; beads mark speech.\\[3pt]
Base code & Communicative move: source-derived label, with unknowns preserved.\\[3pt]
Directed pairing & Attachment: annotated or platform response target; uptake is inspected in text.\\[3pt]
Helix twist & Rhythm: local speaker-switch fraction.\\[3pt]
Helix radius & Reach: source-order target distance.\\[3pt]
Bead radius & Contribution: whitespace word count, with square-root scaling.\\[3pt]
Link texture & Evidence: solid eligible targets versus dashed chronological scaffolds.\\
\bottomrule\end{tabular}
\caption{The visual language separates linguistic interpretation from an observable encoding. Color identifies speakers or moves; apparent depth is only a viewing cue. Numeric definitions appear in Appendix~\ref{app:geometry}.}
\label{tab:grammar}
\end{table}

\section{Implications and Limitations}
Conversation is an achievement of several contributions, not simply a property of any one message. A helpful reply depends on what it takes up; a fluent reply may still address the wrong question. The atlas makes this relational organization available for inspection: whose question receives a response, how clarification branches, and where an apparently continuous exchange changes target. It also makes the observer's role visible. A gap in a strand may be participation structure; a gap in a pairing may be annotation practice. Distinguishing them matters before drawing conclusions about people or comparing human and AI dialogue.

The next evaluation should compare helix, fixed-lane, and transcript views on target tracing, interleaving, and repair-inspection tasks, measuring errors as well as speed. Annotation uncertainty needs separate target and action estimates, validated against human evidence. Multi-turn AI failures \citep{laban2025lost} motivate a further direction: controlled edits to clarification or verification moves, followed by independently measured downstream outcomes. Observed PRISM alternatives do not establish such counterfactual effects. No user study, new annotation-validity study, or causal intervention experiment has yet been completed.

This perspective also changes what an evaluation might ask an AI system to preserve. Beyond producing a plausible next message, can it keep track of an unresolved question, recognize which participant a correction concerns, and respond to a clarification without losing the original task? The atlas can help assemble candidate cases and expose their dependencies. Answering these questions would then require independent judgments of relevance, uptake, and task outcome. For human groups, a similar approach could examine how participants share the work of explanation and repair. Frequent clarification might reflect confusion, careful coordination, or both; its interpretation depends on subsequent contributions. We can connect recurring forms with situated accounts of what participants accomplish, while preserving the evidence that could overturn an appealing visual story.

\section{Conclusion}
Conversational DNA connects a linguistic visual language, an interactive episode atlas, and target-aware retrieval. Its strands and pairings make the organization of contributions inspectable across eight corpora, while source links keep visual readings open to revision. The broader question is how people and AI sustain an exchange together. A useful atlas should help us ask that question more precisely, and recognize when the available evidence cannot yet answer it.

DNA suggests that a conversation’s structure matters through what participants make of it. Future work might draw on transcription and epigenetics to ask how an exchange is retold, which parts become consequential, and how context changes what the same words can do. The deeper question is how people inherit, reinterpret, and transform one another’s contributions into shared meaning.
\label{lastcontent}

\section*{Ethics and Broader Impact}
Public conversation data can contain sensitive material even after filtering. The local application omits WildChat IP hashes, request headers, and geography, and assigns speaker aliases within conversations. These measures do not guarantee that utterance text is anonymous. Raw texts and derived exports remain subject to source licenses, consent conditions, and responsible handling. The companion package's immediate demonstration uses synthetic examples; downloading full corpora is a separate, documented step. No corpus has been newly published as part of this local prototype.

The system is intended for research and evidence inspection. A visual motif or similarity score is not a basis for diagnosing individuals, ranking their character, or making consequential decisions about them. The metaphors of genes, variants, and strands describe representations, not biological properties of speakers. Annotation inventories reflect particular collection settings, and the predominance of English technical and online discussions limits cultural generalization. PRISM ratings represent individual preferences; they should not be interpreted as a population consensus.

AI assistance was used in software implementation and experiment scripting. Reported quantities come from executed code and recorded outputs rather than generated empirical claims. Human author review, verification of attribution, and responsibility for the final submission remain necessary.

\bibliography{references}

\newpage
\appendix
\section{Visual Encoding Specification}
\label{app:geometry}
Let a displayed episode have $n$ turns, $K$ speakers, and speaker index $s\in\{0,\ldots,K-1\}$. Its vertical coordinate is $y_j=58+54j$ pixels (55-pixel steps in the grammar specimen). At each turn $j>0$, the local switching fraction $r_j$ averages whether adjacent speakers differ over available transitions indexed from $j-2$ to $j+2$. Angular advance is
\begin{equation}
\theta_j=\theta_{j-1}+0.42+0.62r_j.
\end{equation}
The initial angle is an adjustable viewpoint. Fixed-pitch mode instead advances by 0.72 radians per turn. With display width $W$, base radius $R=\min(0.22W,132)$, and maximum absolute source-order distance $d_j$ among eligible outgoing targets (zero if unavailable),
\begin{align}
R_j&=R\left(0.72+0.28\min(d_j/6,1)\right),\\
x_{js}&=W/2+R_j\cos\left(\theta_j+\frac{2\pi s}{\max(2,K)}\right).
\end{align}
Fixed-radius mode uses $R_j=R$. Eligible boundary targets contribute to this local reach encoding, but only links with both endpoints visible are drawn. Boundary metadata remain in source evidence. Atlas descriptors count internal episode edges only; their population summaries therefore use a different, declared scope.

Bead radius is $\min(17,7+1.15\sqrt{w_j})$, where $w_j$ counts nonempty whitespace-delimited words. Fixed-size mode uses 11 pixels. All radii are display units, not semantic measurements. Backbone segments interpolate angle and radius. Shading and stroke thickness vary with the apparent sine-depth of the helix for viewing; they carry no uncertainty, confidence, sentiment, or importance variable. A bead belongs only to the speaker active at that turn. The specimen shows at most ten turns and labels its source range; the episode explorer supports up to 24 displayed turns.

\section{Atlas Descriptors and Counting Units}
The 19 descriptors are: log turn count; distinct speakers divided by turn count; adjacent speaker-switch fraction; normalized speaker entropy; largest speaker share; known-action fraction; eligible edges per turn; nonadjacent-edge fraction; mean and maximum target span divided by $n-1$; targets with multiple incoming responses divided by $n$; fractions of clarification, answer, question, acknowledgement, and proposal labels; normalized action entropy; mean log word count; and mean adjacent-turn lexical recurrence. Here word tokens use a lowercase regular-expression word tokenizer, and recurrence is token-set Jaccard overlap, not semantic similarity. Entropies divide by the log of at least two categories. The display grouping retains source labels separately and maps unannotated user/assistant roles to unknown actions for these descriptors.

StandardScaler and full-SVD PCA are fitted to all indexed episodes. Zero target density records target absence in the projection, while the raw cohort statistics retain null span where no target is eligible. Coordinates are rescaled per axis to $[0,1]$ for display. Saved metadata include means, scales, loadings, extrema, variance fractions, corpus order, and episode identifiers. Deterministic ID-dependent jitter of at most 0.006 per axis separates coincident points for selection; brushed membership uses these same displayed positions. Color can instead encode speaker count, response reach, or known-action coverage.

Cohort group share counts episodes containing at least three speakers. Mean target span averages the per-episode means over episodes with eligible links. The nonadjacent fraction pools eligible edge occurrences within the selected episodes. Action coverage averages the per-episode known-action fractions. The census counts target-label to response-label occurrences across windows, including overlap. Consequently it differs from unique-source edge counts. Counts, filters, region bounds, and selected episode IDs can be exported.

\section{Alignment and Evaluation Details}
\label{app:alignment}
Known matching labels score $+2$, differing known labels $-1$, unknown-involving pairs $+0.25$, and gaps $-1$. A dynamic program proposes up to 12 distinct tracebacks from high-scoring endpoints. For alignment $M$, $A(M)$ is its nonnegative sequence score divided by $2|V_Q|$. $S(M)$ is the fraction of aligned node pairs agreeing on whether their respective speakers are identical. With eligible query edges $E_Q$, candidate edges $E_C(M)$ whose endpoints are aligned, and matching edges $T(M)$,
\begin{equation}
F_E(M)=\frac{2|T(M)|}{|E_Q|+|E_C(M)|}.
\end{equation}
Native edge types and both endpoints must match. Equation~\ref{eq:score} is a ranking signal, not a probability. Bounded candidates and tracebacks can miss valid correspondences. Native inventories further limit cross-corpus comparison.

Conversation families join records sharing at least four distinct normalized utterances of at least 20 characters, covering at least 80\% of the smaller qualifying utterance set. Connected components define families; exact normalized full-transcript overlap is also checked. Relevance requires an order-preserving match to all query labels, pairwise speaker relationships, topology, and native edge types. The 2--100-positive eligibility restriction excludes singleton and unseen structures. Query motifs can skip intervening turns; the public browser selects contiguous episodes. The benchmark therefore evaluates the alignment routine on a distinct query workload, not every interface operation.

\section{Reproducibility Details}

The codes to reproduce all the evaluation results and web-based demostration app are included in the GitHub repository: \url{https://github.com/doerlbh/ConversationalDNA}. The portable package to run the application includes clearly labeled synthetic examples. Reproducing the research atlas requires separately obtaining the source datasets subject to their individual licenses.

The retrieval evaluation saves candidate split IDs, each query's source positions and motif key, per-query metrics, and rewiring controls. Family grouping is based only on source text, not labels or retrieval results. The training pool decreases from 9,000 records to \PoolCount{} representatives; the test pool decreases from 500 records to \TestFamilies{} representatives after both cross-split and within-test grouping. Eight retained test families have no eligible motif under the 2--100-positive restriction. The remaining \QueryCount{} families each supply one query. An exhaustive exact-motif lookup provides a 94.2\% P@5 ceiling (some queries have fewer than five positives). The approximate aligner is not optimal for exact motif lookup; it supports longer partial matches using the same representation. No result was used to select the score weights.

TF--IDF uses sublinear term frequency, a minimum document frequency of two, and at most 60,000 word unigram/bigram features. The MiniLM baseline uses Sentence Transformers \texttt{all-MiniLM-L6-v2} through FastEmbed 0.8.1 on CPU, normalized mean-pooled embeddings, cosine similarity, and the backend's 128-token truncation limit. This limit can discard transcript content; the baseline is not an exhaustive comparison of modern text encoders. All structural methods operate on human-provided annotations. The benchmark scores full candidate conversations; the application scores bounded indexed windows. Their latency measurements therefore describe different workloads.

\section{Additional Atlas Screens and Readings}
\label{app:screens}
The following captures document complementary levels of analysis: a corpus-filtered cohort, a source-linked visual grammar, and recorded reply alternatives. They are actual application excerpts. The accompanying retrieval plot summarizes the separate structural benchmark.
\begin{figure*}[t]
\centering\includegraphics[width=\textwidth]{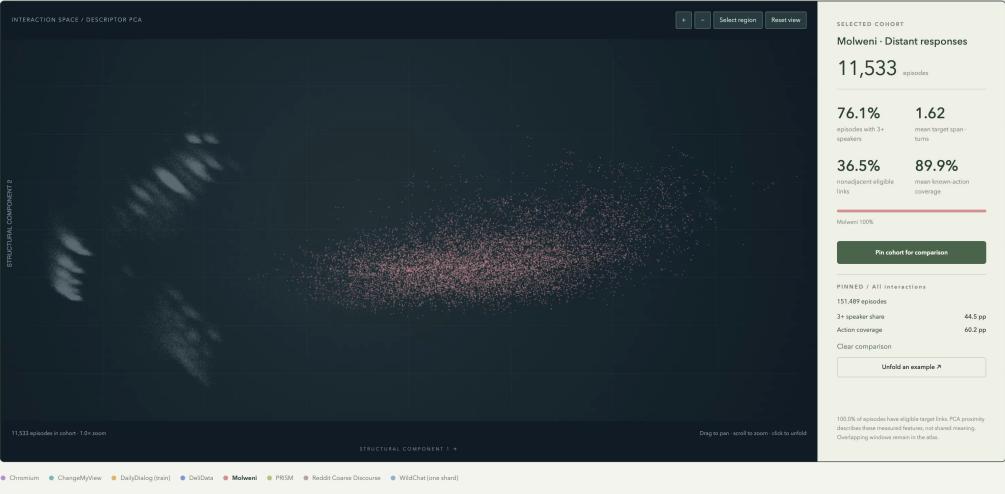}
\caption{A pinned full-population cohort compared with Molweni episodes containing a nonadjacent eligible response. The selected 11,533 episodes have 76.1\% multiparty participation and 89.9\% mean known-action coverage, respectively 44.5 and 60.2 percentage points above the full-map baseline. Showing annotation coverage beside participation makes a potential collection confound inspectable. This is a real application screenshot; no causal comparison is implied.}
\end{figure*}
\begin{figure*}[t]
\centering\includegraphics[width=\textwidth]{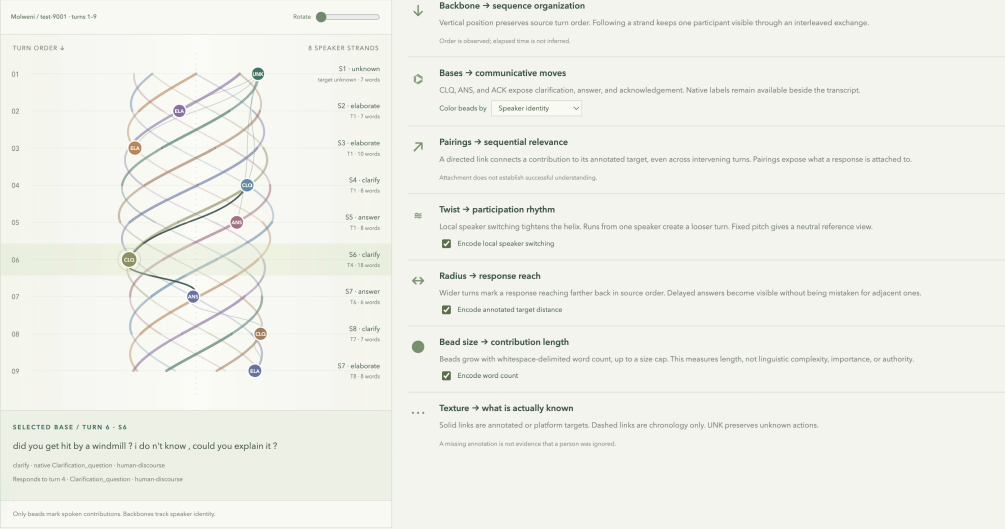}
\caption{The visual-grammar workspace. A source-linked DNA specimen sits beside the seven declared encoding principles. Users can inspect a base, rotate the viewpoint, or disable geometric channels while retaining the same conversation. The selected clarification's original text and native target annotation appear below the specimen. The encodings express observed structure; perceptual advantages over lanes and transcripts have not yet been measured.}
\end{figure*}
\begin{figure*}[t]
\centering\includegraphics[width=\textwidth]{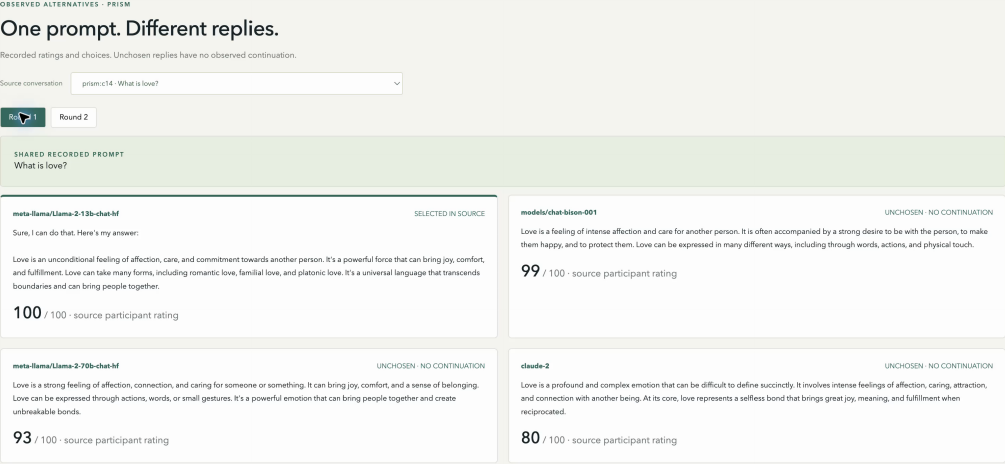}
\caption{Four recorded first-round response alternatives in PRISM conversation \texttt{c14}. Replies at the same source prompt retain model, rating, and selection status. The selected path is observed; unchosen replies have no invented continuation. Ratings are individual participant judgments. This view operationalizes ``variants'' without equating observations with causal counterfactuals.}
\end{figure*}

\begin{figure*}[t]
\centering\includegraphics[width=\textwidth]{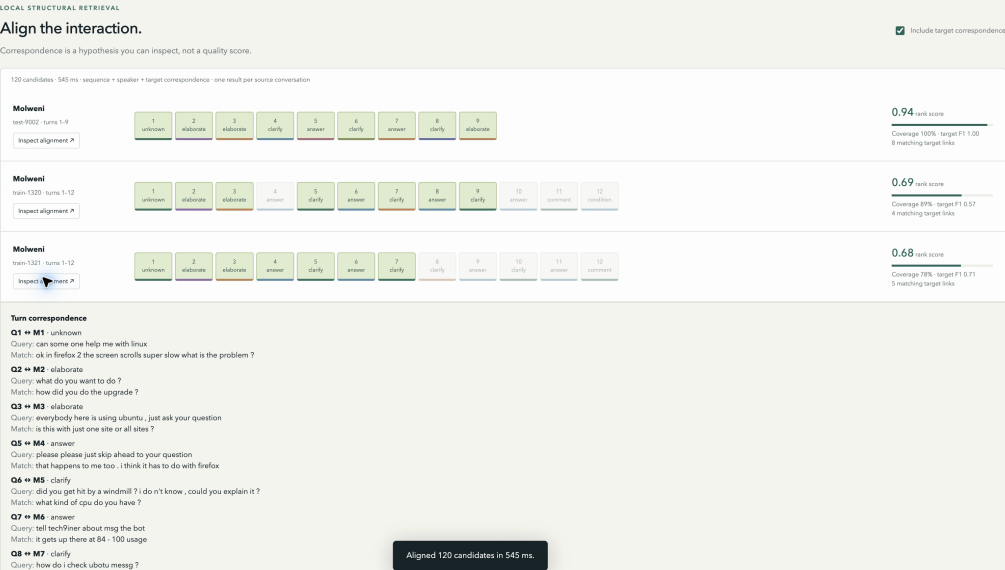}
\caption{Inspecting a retrieved structural analog. The selected Molweni query, \texttt{test-9001}, is aligned with \texttt{train-1321}, an exchange about Firefox. The interface exposes the matched turn positions and original words below the ranking results. Similar interaction structure can connect different topics without making the conversations semantically equivalent. Ranking scores describe the application's current query; they are separate from the held-out benchmark in Table~\ref{tab:retrieval}.}
\end{figure*}

\section{Source Processing and Demonstration}
Chromium's median source record has two utterances; only 54,068 records have at least four. DailyDialog uses its training split and speaker lanes assume alternation. DeliData contributes 14,003 MESSAGE rows, excluding system/submission rows; predecessor links are chronological. All 500 downloaded conversation-level \texttt{performance\_change} values are zero, and the documentation defines change relative to a previous utterance. We do not use that field as a group improvement outcome.

The WildChat subset is the first of 86 release shards. From 37,208 rows, filtering toxicity flags and exact duplicate transcripts retains 36,821. The application omits IP hashes, headers, and geography. PRISM follows a unique chosen response at each round, stopping if selection is ambiguous. Its source contains 265 rounds with two responses marked chosen; no unique continuation is manufactured. Source corpus terms remain separate from the software license.

The walkthrough begins with the full map and a pinned cohort; moves through a populated target--response cell to its DNA; inspects the source and changes an encoding; retrieves and opens a structural alignment; then compares recorded PRISM alternatives and exports evidence. Full research data are separately acquired and evaluated in the paper. The companion code records their provenance and source subsets.

\end{document}